\documentclass[nofootinbib,prl,aps,reprint,amsmath,amssymb]{revtex4-1}
\usepackage{hyperref}
\usepackage{changes}
\usepackage[section]{placeins}
\usepackage[titletoc]{appendix}
\usepackage{xcolor}
\usepackage{dsfont}
\usepackage{bm}
\usepackage{mathtools}
\usepackage{enumitem}
\DeclareMathAlphabet{\mathpzc}{OT1}{pzc}{m}{it}

\usepackage{calc}
\usepackage{accents}

\newcommand{\average}[1]{\left\langle #1 \right\rangle}

\providecommand{\href}[2]{#2}

\usepackage{amsthm}

\def\be{\begin{equation}}
\def\ee{\end{equation}}
\def\bea{\begin{eqnarray}}
\def\eea{\end{eqnarray}}

\def\obs{\mathcal{O}}

\newcommand{\CD}{\mathcal{D}} 
\newcommand{\CQ}{{\mathcal Q}}
\newcommand{\CR}{\mathcal{R}}

\definecolor{MyB}{rgb}{0.1,0.1,1.0}

\def\apjl{{Astrophys. J. Lett.}}

\def\PRD{{{Phys.\ Rev. D.}}}

\def\sovast{{Sov. Astron.}}

\begin{document}
\title{Observational Tests for Distinguishing Classes of Cosmological Models} 

\author{Asta~Heinesen}
\email{asta.heinesen@nbi.ku.dk}
\affiliation{Department of Physics \& Astronomy, Queen Mary University of London, UK}
\affiliation{Niels Bohr Institute, Blegdamsvej 17, DK-2100 Copenhagen, Denmark} 

\author{Timothy~Clifton}
\email{t.clifton@qmul.ac.uk}
\affiliation{Department of Physics \& Astronomy, Queen Mary University of London, UK}

\begin{abstract} 

We investigate observational tests that can be used to distinguish between broad classes of cosmological models. This is achieved using curvature-consistency tests of the Friedmann-Lema\^{i}tre-Robertson-Walker (FLRW) models, which we investigate in two scenarios where they can be violated; (i) when the optical properties of the cosmology deviate from the expectations of FLRW, and also (ii) when the large-scale expansion of the cosmology is different from FLRW. We identify useful ways to determine the properties of these alternative scenarios in terms of the violation of the curvature-consistency tests, and propose a new null test that can be used to isolate cosmologies with non-FLRW observational relations. The characteristic signatures we find can be used, together with the results of recent and upcoming cosmological observations, to probe and/or rule out large classes of cosmological models. This becomes an increasingly important task as the number of proposals in the literature increases, as cosmologists attempt to explain tensions, anomalies, and the dark sector of the Universe. Our approach provides a clear route for telling apart these different proposals, and offers a new opportunity for using precision cosmological data to efficiently discriminate between cosmological models.

\vspace{-10pt}
\phantom{a}


\end{abstract}


\maketitle

\section{Introduction}



Cosmology has long been a subject in which new physics has been proposed and investigated, and in which a constant stream of new models has been considered. This is partly due to the unique range of spatial and temporal scales that are accessible within this field, as well as the huge range of energy scales, but is also driven by phenomenological considerations such as the need for dark energy \cite{amendola2010dark}, apparent anomalies \cite{di2026cracks}, and the recent emergence of tensions between data sets \cite{Schoneberg:2021qvd, Anchordoqui:2021gji, CosmoVerseNetwork:2025alb}. In this environment it is extremely useful to be able to distinguish between competing cosmological scenarios, which is the purpose of this {article}: we seek to investigate and construct observational tests that can be used to isolate and differentiate entire classes of cosmological models.


Our starting point for this will be the \emph{curvature-consistency} tests, in which applicability of the Friedmann-Lema\^{i}tre-Robertson-Walker (FLRW) metric itself is investigated \cite{Clarkson:2007pz,Rasanen:2014mca, FLRWtest1, Montanari:2017yma, FLRWtest3,Dinda:2025svh}.
 {These are falsification tests of the FLRW metric, and thus of its underlying assumptions of homogeneity and isotropy.} 
They are null tests that are independent of both the matter content of the universe and the field equations of the governing theory of gravity. 
They therefore allow all proposals that involve changes to the matter sector or the theory of gravity to be distinguished from those that involve changes to the propagation of light and the geometry of the space-time.


To date, most empirical applications of curvature consistency tests have yielded consistency with FLRW, albeit with some quite wide confidence intervals 
\cite{Montanari:2017yma, Liu:2020pfa, Dinda:2025svh, Millard:2026wnd}. A recent non-parametric investigation utilizing Pantheon+ and BOSS/DESI data has, however, resulted in a mild to moderate inconsistency with FLRW \cite{Koksbang:2026jxb,Koksbang:2026wvh}; see also \cite{Camarena:2025upt} for hints of FLRW violation within a  Lema\^{i}tre-Tolman-Bondi parametrization.
More data, and further investigation, is required to determine if this violation will persist, but should it turn out to be a feature of the Universe in which we live then we will need ways in which we can interpret it, as it would immediately exclude the majority of candidate models for explaining either dark energy or the cosmological tensions \cite{Abdalla:2022yfr}. 
 {To date, curvature consistency tests have not been widely investigated outside of the class of FLRW models from which they were designed, although early studies of some individual model scenarios do exist  \cite{Wiltshire:2009db,Wiegand:2010uh, ref1, ref2}.} 

We will investigate two specific classes of alternative scenarios that we show should be expected to violate FLRW curvature-consistency tests; one based on changes to the optical properties of the space-time \cite{1964SvA.....8...13Z, 1966SvA.....9..671D, 1972ApJ...174L.115D, Dyer:1973zz}, and a second based on changes to the geometry of space-time \cite{Ellis:1987zz, buchertrasanen}.  
Both of these are of potential importance for precision cosmology, with each offering interesting pathways for understanding dark energy and resolving the current tensions in parameter estimates \cite{Heinesen:2020sre, Clifton:2024mdy, Ginat:2026fpo}. This investigation is particularly timely due to the possible first detection of a violation of the curvature-consistency tests, but also due to the prospect of tighter and more reliable test results that are likely to be available from  stage IV surveys in the near future.


We will start by reviewing the key equations for the scenarios under consideration. We will then develop predictions for the expected violations of the curvature-consistency tests in these scenarios, and investigate them in limiting cases of interest. We will end by discussing the implications of our results, including the construction of a new null test, before concluding.

\section{Non-FLRW Cosmological Models}

In this paper we will consider two scenarios that allow for the possibility of novel behavior beyond that which is possible in the usual FLRW approach:

\vspace{10pt}
\noindent
{\bf The Dyer-Roeder Approach:} Here the large-scale kinematical properties of the Universe are given by the usual FLRW equations, but beams of light are permitted to contain less matter than the cosmological average \cite{dyer1973distance, Fleury:2014gha}. The idea is to account for the biasing that might be introduced by masking from regions of high density {of baryonic matter}, which could obscure {the view of} more distant objects. 

In this approach the redshift of light is expected to be given by the usual FLRW relations:
\bea
\label{eq:zprime}
\hspace*{-0.2cm} 1+z = \frac{a_0}{a}  
\qquad {\rm and} \qquad
\frac{ {\rm d} z}{{\rm d} \lambda} =  -E_0  \, H\,  (1+z)^2 \, ,  
\eea 
where $a$ is the scale factor, 
$\lambda$ is an affine parameter  {along past-directed null rays}, $E$ 
is the photon energy, and the subscripts indicates evaluation at the present time, $t=t_0$. Taking the Hubble parameter,  {$H \equiv \dot{a}/a$}, to obey the usual Friedmann equations, 
and neglecting the Weyl focusing, 
the angular diameter distance is then given by solving the following equation:
\be
\label{eq:focuseq}
\frac{{\rm d}^2 d_A }{ {\rm d} \lambda^2} = - 4 \pi G \, \alpha \, E_0^2 (1+z)^2  (\rho + p) \, d_{A}  \, . 
\ee 
Here, we have here generalized the usual approach to allow for fluids with non-zero pressure, $p$. The function $0\leq \alpha \leq 1$ incorporates the effect of the postulated under-density, with $\alpha = 1$ corresponding to the FLRW case and $\alpha = 0$ representing the \emph{empty beam} approximation (in which there is no matter at all along the line of sight). 

From Eqs. \eqref{eq:zprime} and \eqref{eq:focuseq} we can write the equation 
\bea
\label{eq:focuseqz}
 \frac{{\rm d}^2 D }{ {\rm d} z^2} 
&=& -  \frac{1}{H} \frac{\rm{d} H }{\rm{d} z}  \frac{{\rm d} D }{ {\rm d} z}    -    \frac{\mathcal{K_{\rm{DR}}}}{H^2}    \,  D 
\, , 
\eea 
where 
$D \equiv (1+z) d_A$ 
is the 
\emph{comoving distance}, and where we have defined the \emph{effective curvature} function  
\be
\label{eq:DReffcurvature}
\mathcal{K}_{\rm{DR}} \equiv K - 4 \pi G\, (1-\alpha) \, \frac{(\rho + p)}{(1+z)^2} \, , 
\ee 
which reduces to the FLRW spatial curvature $K$ when $\alpha=1$, and which will be useful in what follows. We show in the appendix that $D$ must satisfy
\bea
\label{eq:Dineq}
\hspace{-2pt}  
\left.D \right|_{ \alpha =1} \leq D \leq \left.D\right|_{ \alpha =0}
\quad {\rm and} \quad
\left. \frac{D'}{D}\right|_{\alpha =1}  \!\!\!\!\!\! \leq   \frac{D'}{D}  \leq  \left. \frac{D'}{D}\right|_{\alpha =0} ,
\eea  
where primes denote differentiation with respect to $z$. These inequalities will be used below to provide tests of this scenario in terms of direct observables.

\newpage

\vspace{10pt}
\noindent
{\bf Cosmological Back-Reaction:} The most widely used scheme for quantifying the back-reaction of cosmological structure formation on the large-scale expansion of the Universe is based on the Buchert averaging scheme \cite{Buchert:1999er, Buchert:2001sa, Buchert:2019mvq}. For irrotational dust sources, this approach gives the average expansion rate $H_{\CD}\equiv \dot{a}_{\CD}/a_{\CD}$ of a spatial domain $\CD$ with scale factor $a_{\CD}$ as obeying
\begin{eqnarray}
\label{eq:averagedhamilton}
H_{\CD}^2 &\;=\;& \frac{8\pi G \average{\varrho}}{3}  + \frac{\Lambda}{3} -  \frac{\average{\CR} + {\CQ} }{6} 
\\
\label{eq:averagedraychaudhuri}
3\, (\dot H_{\CD} + H_{\CD}^2) 
 &\;=\;& -\, 4\pi G \average{\varrho} \,+ \,\Lambda + {\CQ}\ ,
\end{eqnarray}
where $\average{\dots} \equiv ( \int_{\CD} \dots dV {)/V_\CD}$ denotes the {average over a domain with volume $V_\CD$}, 
while $\CR$ and ${\CQ}$ are the 3D spatial Ricci-curvature and the kinematical  back-reaction scalar, respectively. 

We use R{\"a}s{\"a}nen's results for angular diameter distances and redshift in this approach \cite{Rasanen:2008be}, which have been shown to work well for light propagation over long distances {in non-trivial test cases} \cite{koksbang2019another,Koksbang:2020zej}. These state that Eqs. \eqref{eq:zprime} and \eqref{eq:focuseq} hold under the substitutions $a \mapsto a_\CD$, $H \mapsto H_{\CD}$, and $\rho \mapsto \average{\varrho}$.
We then derive the following equation for the distance $D$: 
\be
\label{eq:focuseqzBR}
\hspace{-6pt} \frac{{\rm d}^2 D }{ {\rm d} z^2}
= -  \frac{1}{H_{\CD}} \frac{{\rm d} H_{\CD} }{\rm{d} z}  \frac{{\rm d} D }{ {\rm d} z}    -    \frac{\mathcal{K_{\rm BR}}}{H_{\CD}^2}    \,  D \, ,
\ee 
where we have again neglected the effects of shear along the null congruence, set the pressure of the dust fluid to zero, and imposed $\alpha = 1$ (to isolate the back-reaction effect). Similarly to the Dyer-Roeder case, we have again defined an effective curvature function:
\be
\label{eq:BReffcurvature}
\mathcal{K}_{\rm BR} \equiv \frac{ \frac{1}{6} \average{\CR} \, + \, \frac{1}{2}  {\CQ} }{(1+z)^2}   \, , 
\ee 
which again reduces to the FLRW curvature parameter $K$ in an FLRW universe (i.e. when ${\CQ}=0$). 

\section{FLRW consistency tests}
 
The most widely used of these tests was proposed by Clarkson, Basset \& Lu \cite{Clarkson:2007pz}, which starts by defining the following function of expansion and distance:
\be 
\label{eq:O}
\mathcal{O} \equiv \frac{H^2 \left( D' \right)^2  -1}{D^2} \, ,
\ee
and which after differentiation, and multiplication by $\frac{1}{2} D^3/D'$, gives
\begin{align} 
\label{eq:C}
   \mathcal{C}  
   &\equiv 1 + H^2 D  \,D'' - H^2 \left( D'\right)^{2}  + H \, H' \, D  \, D' \, . 
\end{align}
The idea behind defining these quantities is that they reduce to $\mathcal{O}=-K$ and $\mathcal{C}=0$ for \emph{every} FLRW model, independent of matter content or gravitational theory. 

For our current purposes, it is useful to introduce the new test statistic
\begin{align} 
\label{eq:A}
   \mathcal{A}  &\equiv 
   \frac{\mathcal{C}}{D^2}   +  \mathcal{O}  =  H^2  \frac{ D'' }{D}   + H \,H'  \, \frac{D'}{D}\, , 
\end{align} 
which obeys the pleasing identity $D^2 \mathcal{A}' = \mathcal{C}'$, and which also reduces to $\mathcal{A}=-K$ in \emph{every} FLRW cosmology.


We will now consider how the $\mathcal{O}$, $\mathcal{C}$ and $\mathcal{A}$ parameters take values that violate the FLRW expectations of constancy, or disappearance, for the Dyer-Roeder and back-reaction scenarios discussed above.

\vspace{10pt}
\noindent
{\bf Dyer-Roeder induced violations:} In this case we can bound the value of $\mathcal{O}$ by using the inequalities in Eq. \eqref{eq:Dineq}: 
\begin{flalign}
\label{eq:OFLRW_Bound}
&
{\rm (for\:DR)} \hspace{1cm}
- K \leq \mathcal{O} \leq \mathcal{O}\vert_{\alpha = 0} \, .
& 
\end{flalign}  
Thus, there is a sense in which the Dyer-Roeder effect makes the Universe appear more negatively (or less positively) curved when viewed through the value of the $\mathcal{O}$ function. In addition, Eqs. \eqref{eq:focuseqz} and \eqref{eq:DReffcurvature} can be used to show that $\mathcal{A}$ takes the very simple form
\begin{flalign}
&
{\rm (for\:DR)} \hspace{1cm}
\label{eq:A_DR}
   \mathcal{A}  = -\mathcal{K}_{\rm{DR}}   \, . 
& 
\end{flalign}  
A measurement of $\mathcal{A}$, in a Universe in which the Dyer-Roeder approach is valid, therefore provides a direct estimate of the background curvature and the Dyer-Roeder effect through Eq. \eqref{eq:DReffcurvature}. 

The parameters $\mathcal{O}$ and $\mathcal{C}$ are less easily interpretable than $\mathcal{A}$, but at small redshifts Eqs. \eqref{eq:C}, \eqref{eq:A} and \eqref{eq:A_DR} give
\begin{flalign}
&
{\rm (for\:DR)} \hspace{0.5cm}
\label{eq:O_DR_z0}
\mathcal{O}  =   - \left.\mathcal{K}_{\rm{DR}} \right\vert_{t_0}  -  \frac{2}{3}  \left. \mathcal{K}'_{\rm{DR}}\right\vert_{t_0} z +O(z^2)       \,  .
& 
\end{flalign} 
We also have that the following is true for all $z$:
\begin{flalign}
&
{\rm (for\:DR)} \hspace{0.5cm}
\label{eq:Cprime_DR}
\mathcal{C}'     =  - D^2 \mathcal{K}'_{\rm{DR}} \, ,
& 
\end{flalign} 
as follows from directly from $ \mathcal{C}'=D^2 \mathcal{A}'$. The right-hand sides of Eqs. \eqref{eq:O_DR_z0} and \eqref{eq:Cprime_DR} can be evaluated by using Eq. \eqref{eq:DReffcurvature} and the Friedmann equations to obtain
\be 
\label{eq:C_DR_expand_interpret}
\hspace{-0pt}   \mathcal{K}'_{\rm{DR}}  =  -\frac{4 \pi G(1 - \alpha)  (\rho + p) \left( 1 - \frac{\dot p}{H(\rho + p) }  - \frac{(1+z)\alpha'}{1 - \alpha}    \right) }{(1+z)^3}       \, . \,    
\ee
Realistic fluids, which satisfy the null energy condition\footnote{Note that even for evolving dark energy models with phantom properties, which are sometimes considered in efforts of mitigating cosmological tensions, the total cosmological energy budget can still satisfy the null energy condition \cite{qiu2008null}.} ($\rho +p \geq 0$) and have vanishing or negative time-derivatives of pressure ($\dot{p}\leq 0$), therefore have the sign of the initial gradient of $\mathcal{O}$, and the gradient of $\mathcal{C}$ throughout, determined by the rate of change of $\alpha$ with redshift. 

\vspace{10pt}
\noindent
{\it Special case: $\alpha$=constant.} In this case Eq. \eqref{eq:Cprime_DR} reduces to the particularly simple form  
\begin{flalign}
&
{\rm (for\:}\alpha  {\rm \: const}) \hspace{0.35cm}
\label{eq:C_DR_alphaconst}
\mathcal{C}'  = 4\pi G (1 - \alpha)  D^2  \frac{ \left(\rho + p - \frac{1}{H}\dot{p}   \right)}{(1+z)^3}  .  
& 
\end{flalign}
For physical fluids, we thus have that $\mathcal{C}'$ is positive and bounded from above by the empty beam approximation:
\begin{flalign}
&
{\rm (for\:}\alpha  {\rm \: const}) \hspace{1cm} 
\label{eq:C_DR_alphaconst_bound}
0 \leq \mathcal{C}' \leq \mathcal{C}' \vert_{\alpha = 0}  \, .  
& 
\end{flalign}
From $\mathcal{C} = \mathcal{C} \vert_{\alpha = 0} = 0$ at $t_0$, we then have that 
\begin{flalign}
&
{\rm (for\:}\alpha  {\rm \: const}) \hspace{1cm}  
\label{eq:C_DR_alphaconst_bound2}
0 \leq  \mathcal{C}   \leq  \mathcal{C} \vert_{\alpha = 0}  \, .  
& 
\end{flalign} 
We therefore have that constant $\alpha$ very generically implies that $\mathcal{C}$ is non-negative, which directly implies (via Eq. \eqref{eq:C}) that $\mathcal{O}$ is a growing function of redshift. We also have that the empty beam approximation gives an upper bound on $\mathcal{C}$. 

\vspace{10pt}
\noindent
{\bf Back-reaction induced violations:} Using the same methods, we find that the parameter $\mathcal{A}$ takes the following form in the backreaction case: 
\begin{flalign}
&
{\rm (for\:BR)} \hspace{1cm} 
\label{eq:A_BR}
\mathcal{A} =   - \mathcal{K}_{\rm BR} \, .  
&
\end{flalign}
From the definition of $\mathcal{K}_{\rm BR}$ in Eq. \eqref{eq:BReffcurvature}, we see that $\mathcal{A}$ directly measures a linear combination of back-reaction and average spatial curvature in this scenario. Now, using Eqs. \eqref{eq:O} and \eqref{eq:A_BR} we find
\begin{flalign}
&
{\rm (for\:BR)} \hspace{0.5cm}  
\label{eq:O_BR}
\mathcal{O}=  - \mathcal{K}_{\rm BR} \vert_{t_0} - \frac{2}{3} \mathcal{K}'_{\rm BR}\vert_{t_0} z +O(z^2) \, ,
&
\end{flalign}
while we find the following to be true for all $z$:
\begin{flalign}
&
{\rm (for\:BR)} \hspace{0.5cm}  
\label{eq:Cprime_BR}
\mathcal{C}'   =  - D^2   \mathcal{K}'_{\rm BR}  \, . 
&
\end{flalign}
In the back-reaction case we therefore find that the sign of the gradient of $\mathcal{O}$ at the observer, and of $\mathcal{C}$ throughout, is given by the derivative of the linear combination of $\langle \mathcal{R} \rangle$ and $\mathcal{Q}$ given in Eq. \eqref{eq:BReffcurvature}.  

\vspace{10pt}
\noindent
{\it Special case: Scaling solutions.} When back-reaction and curvature follow simple power laws, it can be deduced from Eqs. \eqref{eq:averagedhamilton} and \eqref{eq:averagedraychaudhuri} that\footnote{The excluded case, $n=-2$, results in $\average{\CR}_{\CD} = 6 K (1+z)^2$, which is identical to a spatially-curved FLRW model.}, for $n\neq -2$, \cite{larena2009,Heinesen:2020sre}  
\be 
{\CQ} = \frac{{\CQ}_{0}}{(1+z)^{n}}  \, ,  \quad  \average{\CR} = 6 K (1+z)^2 - \frac{n+6}{n+2} {\CQ} \, ,
\label{eq:scaling}
\ee 
where ${\CQ}_{0}={\CQ}_{0}(t_0)$. Eqs. \eqref{eq:BReffcurvature}, \eqref{eq:A_BR} and \eqref{eq:scaling} then give
\begin{flalign}
&
{\rm (for\:BR)} \hspace{0.4cm}  
\label{eq:A_BRscaling}
\mathcal{A} =   - K - \frac{   \,  \frac{n}{3(n +2)}  {\CQ}_{0} }{(1+z)^{2+n}}   \, .  
&
\end{flalign}
This shows that a scaling index $-2 < n < 0$ with a positive ${\CQ}$ acts as a decaying negative curvature  contribution in Eq. \eqref{eq:A_BRscaling}. Conversely, if $n < -2$ then positive ${\CQ}$ acts as a growing positive curvature contribution. We can use Eqs. \eqref{eq:O_BR} and \eqref{eq:scaling} to get 
\begin{flalign}
&
{\rm (for\:BR)} \hspace{0.4cm}   
\label{eq:O_BR_scaling}
\mathcal{O} =   - K -    \,  \frac{n\, {\CQ}_{0}}{3(n +2)}    + \frac{2n\, {\CQ}_{0}}{9}     \, z+O(z^2) \, . 
&
\end{flalign}
Using \eqref{eq:Cprime_BR} together with \eqref{eq:BReffcurvature}
\begin{flalign}
&
{\rm (for\:BR)} \hspace{0.4cm}    
\label{eq:C_BR_lin_scaling}
\mathcal{C}' =   \frac{n \, D^2\,  {\CQ}_{0}}{3(1+z)^{3+n}}   \, .
&
\end{flalign}
Again, for a scaling index $-2 < n < 0$, positive $\mathcal{Q}$ acts as a negative curvature contribution to $\mathcal{O}$, which decays at higher redshifts, with the contribution to $\mathcal{C}$ being a negative and decaying function. 
\vspace{-5pt}
\section{Discussion}
\vspace{-5pt}
We have shown that the Dyer-Roeder and back-reaction effects can be clearly identified using curvature-consistency tests for $\mathcal{O}$ and $\mathcal{C}$, and we have made predictions for the corresponding signatures that are expected. We have also identified an alternative test statistic, $\mathcal{A}$, which is particularly useful for probing both scenarios, and which we expect to be particularly useful in probing non-FLRW cosmological models more broadly whenever the focusing equation is of the form \eqref{eq:focuseqz} or \eqref{eq:focuseqzBR}. {Our analytic results can be compared to the numerical study of curvature consistency tests that arise from the Dyer-Roeder effect in Ref. \cite{ref1}, and the results that arise from a `template metric' treatment of cosmological back-reaction\footnote{{The `template metric' approach uses an FLRW-like geometry with the same large-scale expansion and spatial curvature as the universal average of the inhomogeneous Universe, but does not reproduce observational relationships such as luminosity distance as reliably as the result of Refs. \cite{Rasanen:2008be} (see e.g. \cite{koksbang2019another,Koksbang:2020zej}). }} in Ref. \cite{ref2}}.

In addition to the test statistic $\mathcal{A}$, we also find that we can construct a new null test statistic for a universe that obeys the Dyer-Roeder equations:
\begin{eqnarray} \label{T}
\mathcal{T} = &&\frac{(1-\alpha) H'}{(1+z)^2 H} - \frac{D'^2 H'}{D^2 H} - \frac{(1-\alpha) H'^2}{(1+z) H^2} \\ \nonumber &&+ \frac{D'H'^2}{D H^2} + \frac{H' \alpha'}{(1+z) H \alpha} - \frac{D' H' \alpha'}{D H \alpha} - \frac{D' D''}{D^2} \\ \nonumber &&+ \frac{3 H' D''}{H D}-\frac{\alpha'D''}{\alpha D }- \frac{(1-\alpha) H''}{(1+z) H} + \frac{D'H''}{DH} +\frac{D'''}{D} \, .
\end{eqnarray}
When $\alpha=1$, so that the usual FLRW equations are recovered, this expression for $\mathcal{T}$ can be seen to be given by the derivative of $\mathcal{C}$. However, unlike any of the test statistics considered above, it vanishes for {\emph{any}} universe obeying the Dyer-Roeder equations, independent of the matter content of the universe or the theory of gravity.

While potentially useful, the $\mathcal{T}$ statistic does come with some drawbacks. Firstly, it is more difficult to apply than the $\mathcal{C}$ statistic, as it contains second derivatives of $H$ and third derivatives of $D$, with respect to redshift $z$. This will undoubtedly mean that it is more difficult to extract from astronomical data. It also requires knowledge of the Dyer-Roeder parameter $\alpha$, and its derivative, which would need to come from either some other set of observations or from {some separate set of} theoretical considerations. 
Nevertheless, it provides a possible way to distinguish the Dyer-Roeder scenario, which would have $\mathcal{C} \neq 0$ and $\mathcal{T}=0$.

\section{Conclusions} 
\vspace{-5pt}
{The results of this paper present direct ways of testing the Dyer-Roeder and back-reaction effects using astrophysical data. }
 {Our proposed tests} provide a way to distinguish these scenarios from the the many alternative proposals for modifying the $\Lambda$CDM model, such as evolving dark energy, interacting fluids, and modified gravity, which would all conform to the standard FLRW expectations of the curvature-consistency tests. Such tests therefore provide a way to falsify these popular ideas, and hence to directly distinguish between different classes of cosmological models.
Of course, violations of curvature-consistency tests may also be induced by other means, and one needs to be confident that they are not polluted by unidentified astrophysical systematics. In order to clearly distinguish between such effects, and from those appearing from a cosmological origin, one can assess whether test violations conform with the theoretical predictions of the Dyer-Roeder effect and back-reaction effects that we have presented here. Such an approach would be bolstered by considering consistency between datasets, as a cosmological signature would be expected to persist across independent catalogs of data, whereas astrophysical systematics are likely to produce internal inconsistencies within the data that they affect only. A further subtlety in using these tests, particularly when considering multiple data sets, is that optical effects (such as those of the Dyer-Roeder equations) could be frequency dependent. {It is also possible that the characteristic signatures of the Dyer-Roeder and cosmological back-reaction effects (as identified above) could be mimicked by some other as-yet-unidentified non-FLRW phenomena of genuine cosmological origin. One must therefore be careful in linking violations of these tests with any one scenario, without corroborating evidence for that effect from elsewhere. However, we expect the existence of quantities such as the $\mathcal{T}$-statistic from Eq. (\ref{T}) to help with this particular problem. We also note that the Dyer-Roeder and cosmological back-reaction effects are special FLRW-violating mechanisms in that they produce isotropic violations of FLRW observational relations. This is in contrast to non-FLRW models with large-scale structures, which otherwise tend to produce directionally-dependent observational features, and which can therefore be used to tell them apart from the scenarios that we have considered in this paper. }

We note that the results we have presented, in combination with current data sets, already have some falsifying power: state-of-the-art symbolic regression and Gaussian process constraints have roughly 10\% uncertainties on the test statistic {$\mathcal{C}$} \cite{Dinda:2025svh,Koksbang:2026wvh}, which means that Dyer-Roeder and back-reaction models with effects of order unity may already be excluded. 
 {An independent investigation of the tests involving {$\mathcal{A}$}, presented in this paper, demonstrates that they are indeed capable of constraining cosmological back-reaction \cite{Koksbang:2026tfw}. The main bottleneck for improving such constraints is the availability of model-independent measurements of $H$ and its derivatives. This will improve in the coming decade with new baryon acoustic oscillation and cosmic chronometers data from e.g. Euclid, which will significantly improve the power of our proposed tests. } 
{Due to the presence of third derivatives, obtaining constraints on the $\mathcal{T}$ statistic is expected to be comparable in difficulty to measuring the cosmic jerk, $j$, which has recently been claimed to be inferred to an accuracy of a couple of percent within the standard cosmological scenario by using fast radio bursts, DESI DR2, and other complementary data \cite{sales2026cosmographic}. Again, this situation is likely to improve further in the coming decades with Stage-IV surveys such as Euclid \cite{blanchard2019euclid}, though the details of exactly what to expect will be strongly dependent on the statistical methods used and the assumptions made (such as degree of smoothness of the relevant functions). We leave a detailed study of such forecasts to future work.}

Finally, we note that curvature-consistency tests are usually examined model-independently, which is at least part of the reason for the relatively loose constraints. However, another possibility is doing a parametrized analysis within particular models that violate the expectations of the test statistics. This could be done within either Dyer-Roeder or back-reaction scenarios, and should be expected to allow tighter constraints from existing data. Calculating predictions for other curvature-consistency tests, such as those presented in Refs. \cite{Rasanen:2014mca,Dinda:2025svh}, would also be very interesting.

\section{Acknowledgments}
We thank Richie L. DeMott for discovering a missing factor of 3 in our calculations for the special case of scaling solutions (Eqs. \eqref{eq:A_BRscaling}--\eqref{eq:C_BR_lin_scaling}), which we have corrected in this updated arXiv version of our paper. 


\appendix 

\section{\label{app:bounds} Appendix: Bounds On Distances And Their Derivatives In The Dyer-Roeder Approach}

Consider the function $\mathcal{W}_c$ and its derivative 
\begin{flalign}
&
\label{eq:dAfun}
\hspace{6pt}  
\mathcal{W}_c \equiv \frac{{\rm d} \left( \frac{ d_{A}\vert_{\alpha =c} }{ d_{A} } \right)}{{\rm d} \lambda}  =    \frac{{\rm d} d_{A}\vert_{\alpha =c} }{ {\rm d} \lambda} d_A -   \frac{{\rm d} d_A }{ {\rm d} \lambda}  d_{A}\vert_{\alpha =c} 
&
\\[6pt]
&
\nonumber
\hspace{6pt}  
\frac{ {\rm d} \mathcal{W}_c }{  {\rm d} \lambda }  =   \frac{{\rm d}^2 d_{A}\vert_{\alpha =c} }{ {\rm d} \lambda^2} d_A -   \frac{{\rm d}^2 d_A }{ {\rm d} \lambda^2}  d_{A}\vert_{ \alpha =c} \,  \begin{cases}
    \geq 0 & \text{if $c = 0$} \\
    \leq 0 & \text{if $c = 1$}
  \end{cases}
  &
\end{flalign} 
where the inequalities follow from Eq. \eqref{eq:focuseq}. This implies 
\bea
\label{eq:dAfun_bound}
\hspace{-6pt}  
\mathcal{W}_c =  \begin{cases}
    \leq 0 & \text{if $c = 0$} \\
    \geq 0 & \text{if $c = 1$}
  \end{cases}
\eea 
for all $\lambda \leq \lambda_\obs$ (i.e. on the observer's past light cone), where we have used the initial condition $\mathcal{W}_{c}(t_0)  = 0$. In an expanding universe, with $H > 0$, the above inequalities can be combined with Eq. \eqref{eq:zprime} to find that
\bea
\label{eq:dAineq_z}
\hspace{-6pt}  
 d_A \leq d_{A}\vert_{\alpha =0} \, , \quad \frac{\frac{{\rm d} d_A }{ {\rm d} z}}{d_A}  \leq \left. \frac{\frac{{\rm d} d_{A} }{ {\rm d} z}}{d_{A}} \right\vert_{\alpha=0} 
\eea 
and 
\bea
\label{eq:dAineq2_z}
 d_A \geq d_{A}\vert_{\alpha =1} \, , \quad \frac{\frac{{\rm d} d_A }{ {\rm d} z}}{d_A}  \geq \left. \frac{\frac{{\rm d} d_{A} }{ {\rm d} z}}{d_{A}} \right\vert_{\alpha=1}   
\eea
for all values of $z$. Finally, rewriting Eqs. \eqref{eq:dAineq_z} and \eqref{eq:dAineq2_z} in terms of $D$, we have 
\bea
\label{eq:Dineqapp}
\hspace{-6pt}  
D \leq D \vert_{ \alpha =0} \, , \quad \frac{\frac{{\rm d} D }{ {\rm d} z}}{D}  \leq \left. \frac{\frac{{\rm d} D }{ {\rm d} z}}{D} \right\vert_{\alpha=0}
\,  
\eea 
and 
\bea
\label{eq:Dineq2app}
D \geq D \vert_{ \alpha =1} \, , \quad \frac{\frac{{\rm d} D }{ {\rm d} z}}{D}  \geq \left. \frac{\frac{{\rm d} D }{ {\rm d} z}}{D} \right\vert_{\alpha=1}
\,    \, . 
\eea

\newpage

\bibliography{refs}

\end{document}